# Cryptanalysis of "Cryptanalysis and Improvement of Yan et al.'s Biometric-Based Authentication Scheme for TMIS"

Mrudula Sarvabhatla
Sri Venkateswara University
Tirupathi-517518, A.P.
Mrudula.s911@gmail.com

M.Giri
Professor and H.O.D
SITAMS, Chitoor, A.P
prof.m.giri@gmail.com

Chandra Sekhar Vorugunti
Dhirubhai Ambani Institute of ICT
Gandhinagar-Gujarat.
vorugunti_chandra_sekhar@daiict.ac.in

*Abstract*— Remote user authentication is critical requirement in Telecare Medicine Information System (TMIS) to protect the patient personal details, security and integrity of the critical medical records of the patient as the patient data is transmitted over insecure public communication channel called Internet. In 2013, Yan proposed a biometric based remote user authentication scheme and claimed that his scheme is secure. Recently, Dheerendra et al. demonstrated some drawbacks in Yan's scheme and proposed an improved scheme to erase the drawbacks of Yan's scheme. We analyze Dheerendra et al.'s scheme and identify that their scheme is vulnerable to off-line identity guessing attack, and on successfully mounting it, the attacker can perfom all major cryptographic attacks.

## I. INTRODUCTION

The rapid development in network and communication technology has presented a scalable platform for Telecare Medicine Information System (TMIS). The communication between the user and server is always a subject of security and privacy risk in TMIS as user accesses remote server via public channel and an adversary is considered to be enough powerful to perform various attacks. Thus the secure and efficient authenticated key agreement schemes should be adopted to ensure security and integrity of transmitting data [1]. The smart card based authentication scheme pro-vides efficient solution for remote user authentication [6,7]. In recent times, many password based authentication schemes have been proposed for TMIS [9,10]. These schemes try to provide two factor authentication.

The password cannot be considered as a unique identity identifier and it's needed to be remembered. Moreover, possibility of password guessing attack is also a concern. However, biometrics cannot be lost or forgotten, have the merits of uniqueness and need not be remembered; but they can be compromised [8]. Additionally, these bio-metric keys are not easy to guess [11,12]. Due to these advantages, the biometrics based authentication schemes present efficient solution to mutually authenticate and ses-sion key agreement. In 2013, Tan [1] presented a biometric based remote user authentication scheme for the Telecare medical information system. In Tan's scheme, a remote user and server can mutually authenticate each other and draw a session key. Moreover, the Tan's scheme presents a user-friendly password and biometric update phase where a user can change his password and biometric keys without server assistance. Recently, Yan et al.'s [2] pointed out that Tan's scheme is vulnerable to denial-of-service attack. Further, they proposed an improved scheme to eliminate the draw-backs of Tan's scheme. Their scheme also preserves all the merits of Tan's scheme.

In this article, we analyze the Yan et al.'s biometrics based remote user authentication scheme for TMIS. We show that Yan et al.'s scheme login phase is inefficient such that the smart card executes the login session in-spite of incorrect input. The inefficiency of the login phase in incorrect input detection causes extra communication and computation overhead. Yan et al.'s password and biometrics update phase is also inefficient to detect incorrect input, which causes denial of service attack in case of wrong password input. Yan et al.'s scheme does not withstand pass-word guessing attacks. Furthermore, we present a modified scheme which overcomes the weaknesses of Yan et al.'s scheme and preserves its merits.

The remaining part of the article is organized as follows: Section "Review of Dheerendra et al.'s scheme" presents a brief review of Dheerendra et al.'s scheme. Section "Weaknesses of Dheerendra et al.'s scheme" demonstrates the weaknesses of Dheerendra et al.'s scheme. The conclusion is drawn in section "Conclusion".

The remaining fragment of the paper is structured as follows. In fragment II a brief analysis of Dheerendra et al scheme is given. Fragment III explain the security flaws of Dheerendra et al. scheme and fragment IV gives the conclusion of the paper.

## II. ANALYSIS OF DHEERENDRA ET AL SCHEME

In this part, we inspect the improvement of Dheerendra et al. [3] authentication scheme for TMIS. The scheme is a collection of three phases: the registration, login, authentication and password and biometrics update stage.

## A. Registration Phase

This stage is a one time execution process, when user $U_i$ wish to list with the remote system.

Step 1. $U_i$ selects an identity $ID_i$ and secret password $PW_i$ of his choice, and imprint his biometrics $B_i$. He/She generates a random number $b_i$, and computes $W_i = h(ID_i \| PW_i \| r_i)$. $U_i$ submits the registration request with $ID_i$ and $W_i$ to S via secure channel.

Step 2. S computes $X_i = h(ID_i \| x)$, $Y_i = X_i \oplus W_i$, where x is the server's 1024-bits or 2048-bits secret key. S generates a random number R and computes user's dynamic identity by encrypting the user identity using symmetric key encryption algorithm such as AES-256, i.e., $NID = E_x(ID_i \| R)$. The server selects the long key to resist server's secret key guessing attack. Then S embeds $\{NID_i, Y_i, h(\cdot)\}$ into the smart card and issues the smart card to $U_i$. Step 3. Upon receiving the smart card, $U_i$ stores $N = r_i \oplus H(B_i)$ and $V_i = h(ID_i \| PW_i \| r_i)$ into the smart card.

## B. Login and Authentication Phase

At any time the user in need to access the far-off server S, the subsequent procedure is made. (L1) $U_i$ inputs $ID_i$ and $PW_i$, and imprints his biometrics $B_i$ at the sensor. The smart card computes $N_i = N \oplus H(B_i)$, and verifies $V_i = h(ID_i \| PW_i \| r_i)$ if the verification does not hold, the smart card terminates the session. The smart card computes $W_i = h(ID_i \| PW_i \| r_u)$ to achieve $X_i = Y_i \oplus W_i$. S.C generates a random $a_i = h(ID_i \| X_i \| r_u)$. Then sends the login message $< NID, a_i, r_u>$ to S.

## C. Validation Phase

On intercepting $U_i$'s login request message at time $T^*$, the server S executes the subsequent steps:

(V1) S retrieves $ID_i$ by decrypting NID and computes $X_i = h(ID_i \| x)$. S verifies $a_i$ equal to $h(ID_i \| X_i \| r_u)$. If the verification does not hold, S terminates the session.

(V2) S generates random numbers $r_s$ and $R^*$, and computes $SK=h(ID_i \| X_i \| r_u \| r_s)$, $NID^* = E_x(ID_i \| R^*)$ and $B_i = h(ID_i \| NID \| SK \| NID^*)$. S sends the message $<r_s, B_i, h(SK \| ID_i) \oplus NID^*>$ to the user.

(V3) On receiving the login reply message message $<r_s, B_i, h(SK \| ID_i) \oplus NID^*>$ from S, the SC computes the session key $S.K = h(ID_i \| X_i \| r_u \| r_s)$, and retrieves $NID^* = (SK \| ID_i) \oplus NID^* \oplus (SK \| ID_i)$.

S.C computes $B_i^* = h(ID_i \| NID \| SK \| NID^*)$ and compares with $B_i$, if both are equal S.C authenticates S else rejects the login request. On authenticating the server, S.C computes $C_i = h(ID_i \| NID^* \| S.K)$ and the session key verification message to S.

(V4) On receiving the session key reply message, S computes $C_i^* = h(ID_i \| NID^* \| S.K)$ and compares with the received $C_i$, if both are equal, S fully authenicates $U_i$.

## III. CRYPTANALYSIS OF DHEERAJ ET AL SCHEME

In this segment, we will cryptanalysis the Dheerendra et al scheme and show that Dheerendra et al's authentication scheme is insecure against offline Identity guessing attack and on successful mounting Identity attack, the attacker can perform all major cryptographic attacks

### A. Through stolen smart card of legitimate user:

A legal adversary 'E', if gets the smart card of a valid user $U_i$ of the system for a while or stolen the card, 'E' can extract the secret data stored in $U_i$'s smart card as discussed in [4, 5]. In Dheerendra et al scheme, as discussed in registration stage, 'E' can get $\{NID, Y_i, h(\cdot), N, V_i\}$ which are stored in the $U_i$ smart card, Where $W = h(ID_i \| PW_i \| N_i)$, $X_i = h(ID_i \| x)$, $Y_i = X_i \oplus W_i$, $V_i = h(ID_i \| PW_i \| N_i)$ which means $V_i$ equal to $W_i$.

$Y_i$ is available to the attacker as it is stored in the $U_i$ smart card. Now 'E' can proceed as follows:

$V_i = W_i = h(ID_i \| PW_i \| r_i)$     (1)
$Y_i = X_i \oplus W_i$     (2)
From (1) $Y_i = X_i \oplus V_i$     (3)
From (3) E can intercept $X_i = Y_i \oplus V_i$     (4)

### B. Through intermediate messages exchanged between legitimate user and the server S:

Once legal user $U_i$ logs into the system, the legal adversary 'E' can capture the intermediate login request, login reply messages exchanged between the user and the server S. In Dheerendra et al scheme, the adversary can capture login request $\{NID, a_i, r_i\}$ exchanged between $U_i$ and the server S.

In below subsections, we discuss how Dheerendra et al scheme is vulnerable when an adversary is provided with one or more set of above discussed values.

### C. Failure to Offline Identity Guessing Attack

The identity of a patient is often known to all in the TMIS system. The users usually choose easy recollective names as identity like social security ID, email, phone number and so on as their identities. In authentication and key agreement phase, the user need to input his identity and password to login the server. Even the user intends to keep his identity in secret, however, a easy-to-remember identity for the user is also easy-to-guess for an attacker. Assume that identity is selected from a limited set of uniformly distributed dictionary, then the adversary can proceed as follows :

When the patient $U_i$ logs in to the system and sends the login message $\{NID, a_i, r_i\}$ to S, where $NID = E_x(ID_i \| R)$, the attacker records it as it is transferred through a public channel. In the login request sent by $U_i$, $a_i = h(ID_i \| X_i \| r_u)$. Among $ID_i, X_i, r_u$ the attacker intercepted $X_i, r_u$ as discussed above. Only unknown value is $ID_i$. The attacker can proceed as follows to get the identity of $U_i$.

Step 1 : The attacker selects a candidate identity ID as $ID^*$ from a limited set of uniformly distributed dictionary and computes $a_i^* = h(ID_i^* \| X_i \| r_u)$.

Step 2: Check $a_i^*$ equals $a_i$, If both are equal then the $U_i$ identity is $ID_i^*$, else proceed to step 1 until the correct identiy is found .

On successfully getting the identity $ID_i$ of $U_i$, the attacker can proceed with following attacks:

### D. Failure to resist user Impersonation attack

In user impersonation attack, the adversary 'E' can impersonate as a valid user $U_i$ by forging the login message contents. In Dheerendra et al., scheme a valid user $U_i$ sends the login message i.e., $<NID, a_i, r_u>$ where $NID = E_x(ID_i \| R)$, $a_i = h(ID_i \| X_i \| r_u)$. The adversary 'E' can perform the impersonate attack, when $U_i$ logged into the system as follows.

Step 1: $NID = E_x(ID_i \| R)$, is static entity which doesn't changes with each login, So only value the attacker needs to modify is $a_i = h(ID_i \| X_i \| r_u)$, attacker knows $ID_i$, $X_i$, $r_u$. ($ID_i$ from (C), $X_i$ from (A), $r_u$ from (B) ).

Step 2: To frame a valid login request the attacker can modify $a_i$ by chosing a new random number i.e $a_i^* = h(ID_i \| X_i \| r_u^*)$ and sending $<NID, a_i^*, r_u^*>$ . The login message will sure pass the checks made by server S.

Therefore, we can conclude that in Dheerendra et al. scheme, the adversary can impersonate as a valid user $U_i$, by replaying the previously intercepted authentication messages as discussed above. Hence, Dheerendra et al. scheme is vulnerable to user impersonation and replay attacks.

### E. Attacker can frame the session Key

An attacker can frame the session key framed between $U_i$ and S as follows:

An attacker can intercept the login reply message from S to $U_i$ i.e $<r_s, B_i, M_i>$, where $SK=h(ID_i \| X_i \| r_u \| r_s)$, $NID^* = E_x(ID_i \| R^*)$ and $B_i = h(ID_i \| NID \| SK \| NID^*)$, $M_i = h(SK \| ID_i) \oplus NID^*$.

'E' knows all the parameters to compute session key i.e $ID_i, X_i, r_u, r_s$. ('E' can get $r_s$ from login reply message).

Step 1: As discussed above , 'E' knows $ID_i$, NID, $X_i$, $r_u$, $r_s$. 'E' can frame $S.K = h(ID_i \| X_i \| r_u \| r_s)$,

Hence, 'E' can decrypt all the messages exchanged between $U_i$ and S.

Therefore, we can conclude that in Dheeraj et al. scheme, the adversary can frame the session key and read all the messages exchanged between $U_i$ and S. Hence, Dheeraj et al scheme failed to satisfy the fundamental requirement of the remote user authentication scheme i.e data security.

Therefore, we can conclude that in Dheerendra et al. scheme, Once the identity of $U_i$ is known to the adversary 'E' , he can impersonate the user and frame the session key.

## IV. CONCLUSION

The present paper analyzed the security vulnerabilities in Dheeraj et al biometric based remote user authentication scheme. We have shown that that if an adversary gets the identity of the legal user, then he can frame the session key. In future, we will propose our improved scheme which fixes the vulnerabilities found in Dherendra et al and other related schemes.

## REFERENCES

[1] Z.Tan, "An efficient biometrics-based authentication scheme for telecare medicine information systems" , Network 2(3):200–204, 2013.

[2] X.Yan, W.Li, P.Li, J. Wang, X.Hao, and P.Gong, "A secure biometrics-based authentication scheme for telecare medicine information systems.", springer journal of Medical Systemss 37(5):1–6, 2013.

[3] D.Mishra, S.Mukhopadhyay, A.Chaturvedi, S.Kumari, and M.K.Khan, "Cryptanalysis and Improvement of Yan et al.'s Biometric-Based Authentication Scheme for Telecare Medicine Information Systems", springer ournal of Medical Systems, June 2014.

[4] P.Kocher,J. Jaffe, and B.Jun, "Differential power analysis. In: Advances in Cryptology" , CRYPTO99: Springer, 388–397, 1999.

[5] T.S Messerges, E.A Dabbish, and R.H Sloan, "Examining smart-card security under the threat of power analysis attacks" , IEEE Trans. Comput. 51(5):541–552, 2002.

[6] Khan, M.K., and Kumari, S., An authentication scheme for secure access to healthcare services. J. Med. Syst. 37(4):1–12, 2012

[7] .Kumari, S., Khan, M.K., and Kumar, R., Cryptanalysis and improvement of a privacy enhanced scheme for telecare medical information systems. J. Med. Syst. 37(4):1–11, 2012.

[8] M.K Khan, J. Zhang, and K.Alghathbar, "Challenge-response-based biometric image scrambling for secure personal identification" , Futur. Gener. Comput. Syst. 27(4):411–418, 2011.

[9] T.Cao, and J. Zhai, " Improved dynamic ID-based authentication scheme for telecare medical information systems" , J. Med. Syst. 37(2):1–7, 2013.

[10] H.M Chen, J.W. Lo, and C.K Yeh, "An efficient and secure dynamic ID-based authentication scheme for telecare medical information systems" J. Med. Syst. 36(6):3907–3915, 2012.

[11] M.K Khan, J. Zhang, K. and Alghathbar, "Challenge-response-based biometric image scrambling for secure personal identification" , Futur. Gener. Comput. Syst. 27(4):411–418, 2011

[12] M.K Khan, J. Zhang,, and L.Tian, " Protecting biometric data for personal identification", In: Advances in Biometric Person Authentication: Springer, 629–638, 2005.